\documentclass[a4paper,aps,pre,twocolumn,groupedaddress,showkeys,showpacs, 
  superscriptaddress,floats,floatats,floatfix]{revtex4-1}
\usepackage[latin1]{inputenc}
\usepackage{graphicx}  % Include figure files
\usepackage{dcolumn}   % Align table columns on decimal point
\usepackage{bm}        % bold math
\usepackage{natbib}
\usepackage{latexsym}
\usepackage{mathrsfs}
\usepackage{amssymb}
\usepackage{amsmath}
\usepackage{amscd}
\usepackage{color}
\usepackage{verbatim}
\begin{document}

\title{Intrinsic stickiness in open integrable billiards: tiny border effects}
\author{M.~S.~Cust\'odio and M.~W.~Beims}
\email[E-mail address:~]{mbeims@fisica.ufpr.br}
%\thanks{}
%\altaffiliation{}brasil-argentina.jpg
\affiliation{Departamento de F\'\i sica, Universidade Federal do Paran\'a,
         81531-990 Curitiba, PR, Brazil}
\date{\today}

\begin{abstract}
Rounding border effects at the escape point of open integrable billiards are
analyzed via the escape times statistics and emission angles. The model is the 
rectangular billiard and the shape of the escape point is assumed to have a
semicircular form. Stickiness and self-similar structures for the escape times 
and emission angles are generated inside ``backgammon''  like stripes of initial
conditions. These stripes  are born at the boundary between two 
different emission angles but same escape times. As the 
rounding effects increase, backgammon stripes start to overlap and the escape 
times statistics  obeys the power law decay and anomalous diffusion is 
expected. Tiny rounded borders (around $0.1\%$ from the whole billiard size) 
are 
shown to be sufficient to generate the sticky motion, while borders larger 
than $10\%$ are enough to produce escape times with chaotic decay.
\end{abstract}



\maketitle

\section{Introduction}
\label{Introduction}

Experiments usually measure a signal (light, atoms, particles current etc) 
which comes out from the system of interest. If the open set (the escape point 
or hole) of the physical device, 
where the signal comes out, has a relevant size and shape, the 
signal may include informations from inside the physical device 
{\it and} from  the open set itself. The shape and size of the open 
set depends on experimental interests but also on how the device is build.
For the dynamics in mesoscopic systems and nanostructures, for example, 
the shape and the size
of the small open set can present irregularities or defects which may induce
undesirable changes in the outcoming signal. Such sometimes
intrinsic irregularities affect the dynamics of particles which collide 
with them. 

From the theoretical point of view it is very difficult 
to describe, in general, the dynamics of colliding particles with irregular 
boundaries. Therefore in recent years more and more attention has been given 
to the description of particles confined inside boundaries (or billiards) 
which present some {\it specific} edges, softness etc.
To mention some examples we have the edge roughness in quantum dots 
\cite{libisch09}, unusual boundary conditions in two-dimensional billiards 
\cite{bogo09,dennis08}, effects of soft walls \cite{hercules08}  
and edge collisions \cite{cesar1} of interacting particles in a 1D 
billiard, rounding edge \cite{wiersig03} and edge corrections 
\cite{blumel05} in a resonator, deformation of dielectric cavities
\cite{bogo08},  edge diffractions and the corresponding semiclassical 
quantization \cite{whelan96,gaspard94}, among others. 

Different from the above works, which focus on the boundaries of 
systems, here we analyze the effect of ir\-re\-gu\-la\-ri\-ties from the 
open set itself. What is the effect of (rounded) open sets on the outcoming 
signal ? To mention some examples, rounded open sets are common 
in experiments with semiconductors devices \cite{marlow06,sachrajda98} and 
quantum cavities \cite{takagaki00}, where the open sets have a shape 
very similar to those shown in Fig.~\ref{Model-Border}, which is the model 
used here (Type I and II). In the above mentioned experiments the ratio between 
the open set (not its width, but the radius of the rounded border) and the 
whole device lies  around $\sim 0.001$ for Type II borders (the smallest 
case) and around $\sim0.3$ for Type I borders (the largest case).
Since experiments measure a signal which comes out from the system 
of interest, how is it affected by such rounded borders ? is it possible that 
tiny rounded borders transform an integrable dynamics 
into a chaotic one ? what is their influence on the dynamics in open 
nanodevices, conduction fluctuations in semiconductors 
\cite{geisel07,marlow06}, particles transport in 
nanostructures, cold atoms in open optical billiards  \cite{kaplan09} etc?
Recent works \cite{altmann09,dettmann09, buni07, ott04} in this direction 
analyzed the effect of the 
{\it width} of the open set on the escape rates of particles in open billiards.

Using the example of a billiard model we show here that the 
{\it shape} or {\it structure} of the open set induces stickiness 
\cite{zas02}, self-similar and a chaotic output signal, even if the 
dynamics inside the billiard is originally re\-gu\-lar. 
We consider that the device borders (or extremities), which 
delimit the holes in realistic open systems, are not single points, but have 
their own shape. In this work results are shown for the escape times (ETs) 
statistics and the emission angles $\theta_f$ in the open rectangular billiard 
shown in Fig.~\ref{Model-Border}.
While self-similar structures are clearly visible in  the ETs and
escape angles, the sticky motion is observed by the power law decay of 
the ETs statistic. Although it is possible to detect and quantify sticky 
motion via the distribution of finite time Lyapunov exponents 
\cite{steven07,cesar1,hercules08,cesarPRE09,manchein09}, for the purpose of 
the present work it is more adequate to use the ETs statistics. 
The fractal behavior of the ETs dynamics was also shown recently in 
Bose-Einstein condensates \cite{mitchell09}, trapped ultracold atoms 
\cite{mitchell07} and in a vase-shaped cavity \cite{delos06}, 
in generic chaotic cavities \cite{ketzmerick96}, just to mention 
some examples.

The paper is organized as follows. Section~\ref{model} presents the model,
defines the ETs used to detect the sticky behavior and shows related numerical
results. Sections~\ref{dynamics} and \ref{backgammon} show the rich 
dynamics generated by the rounded border: long-lived states and self-similar
structures for the ETs and emission angles. In Section~\ref{conclusions}
we present our final remarks.

\section{Rounding borders generating stickiness}
\label{model}

The ETs statistic is defined \cite{altmann05} by
$Q(\tau) = \lim_{N\to\infty}\frac{N_{\tau}}{N},$
where $N$ is the total number of trajectories which escape the billiard and
$N_{\tau}$ is the number of trajectories which escape the billiard after the
time $\tau$. For systems with stickiness the ETs statistic decays as a power 
law \cite{altmann05} $Q(\tau)\propto \tau^{-\gamma_{esc}}$, where $\gamma_{esc}>1$ 
is the scaling exponent. For hyperbolic chaotic systems and long times the 
ETs statistic decays exponentially. It is known that the diffusion exponent 
$\mu$ from the mean square displacement of the position 
$\left< x^2\right>\sim t^{\mu}$, is related to $\gamma_{esc}$ via 
 $\mu=3-\gamma_{esc}$ \cite{meiss97,klafter99}.
%%%%%%%%%%%%%%%%%%%%%%%%%%%%%%%%%%%%%%%%%%%%%%%%%%%%%%%%%%%%%%%%
 \begin{figure}[htb]
 \unitlength 1mm
 \begin{center}
 \includegraphics*[width=6cm,angle=0]{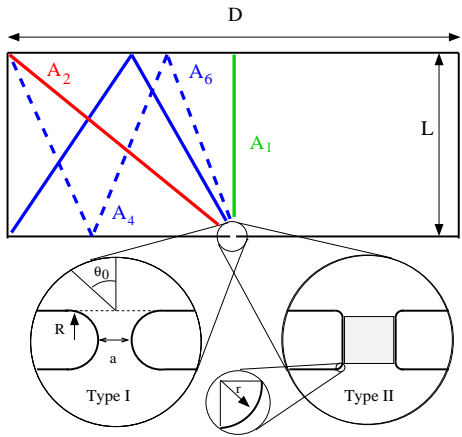}
 \end{center}
\caption{(Color online) The rectangular billiard, with dimension
$L\times D$, showing the open sets (borders) of Type I  (with radius $R$) 
and Type II (radius 
$r\approx R/20$) considerer here. The escape point lies  exactly in the 
middle of the billiard and has constant aperture $a=1\times 10^{-4}$. 
Initial angle $\theta_0$ and, schematically,  the shortest escape trajectories 
($A_1,A_2,A_4$ and $A_6$) are shown. In all simulations we use $L=4$ and $D=10$.}
  \label{Model-Border}
  \end{figure}
%%%%%%%%%%%%%%%%%%%%%%%%%%%%%%%%%%%%%%%%%%%%%%%%%%%%%%%%%%%%%%%%%%%%%

In the simulations, particles start at times $t=0$ from the escape point with 
an initial angle $\theta_0$ towards the inner part of the billiard with 
velocity $|\vec v|=1$. The particle suffers elastic collisions at the
billiard boundaries and at the rounded border of the open set.
For each initial condition we wait until the particle leaves the billiard 
and record $\theta_f$ and the ETs. We use
$10^5$ initial conditions distributed uniformly in the interval 
$0.10\lesssim\theta_0\lesssim 1.54$.  The dynamics for  
$0.0\le\theta_0\le -\pi/2$ is symmetric.
Without rounding effects ($R=r=0$) the rectangular billiard is integrable and 
has zero Lyapunov exponents. As the ratio $R/L$ increases, the rounded border
acts like a dispersing boundary generating chaotic motion without re\-gu\-lar 
islands. For larger values of $R/L$ and some specific initial conditions, the 
dynamics should be equivalent to those obtained in the Sinai 
\cite{sinai} and Bunimovich stadium \cite{buni79} billiards. These billiard 
have marginally unstable periodic orbits (MUPOs), which are orbits bouncing
perpendicularly between the parallel walls, and are known 
\cite{ott04,dorfman95} to generate the sticky exponent $\gamma_{esc}=2$ for
the ETs. In the rectangular billiard the MUPOs are obtained from the initial 
conditions $\theta=0,\pi/2$. Therefore, in our case we also expect to see 
$\gamma_{esc}\to2$ as $R/L$ increases.

\begin{figure}[htb]
 \unitlength 1mm
 \begin{center}
 \includegraphics*[width=8cm,angle=0]{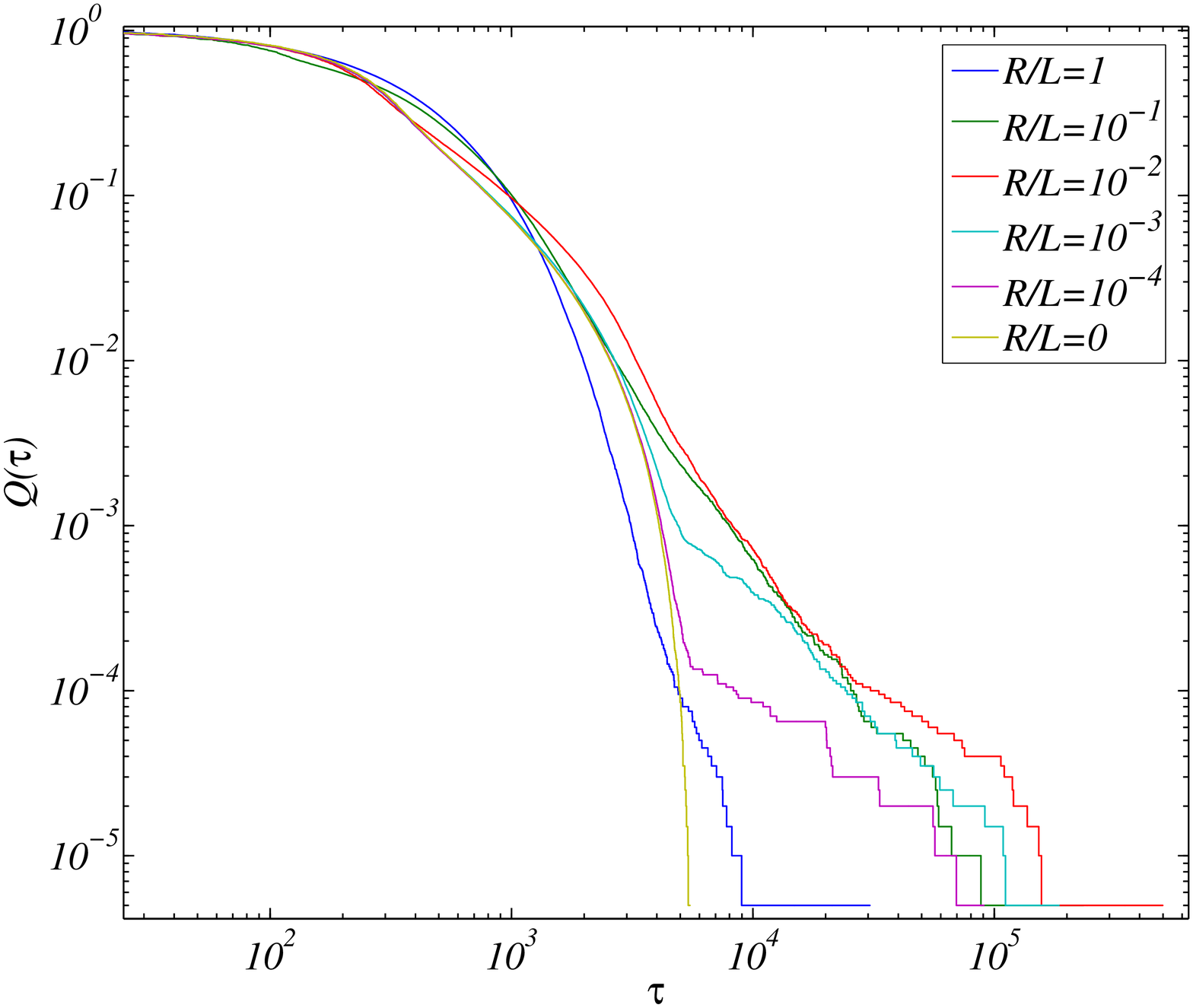}
 \end{center}
\caption{(Color online) Behavior of $Q(\tau)$ for different values
of the ratio $R/L$.}
  \label{QtR}
  \end{figure}

We start discussing numerically the quantity $Q(\tau)$ for the Type I billiard 
and for different values of the ratio $R/L$, where $L$ is kept fixed.  
Results are shown in Fig.~\ref{QtR} for 
$R/L=0, 1/10000, 1/1000, 1/100, 1/10$ and $1$. Note that these results are
automatically applied to the Type II billiard with
$r/L\sim 0, 1/200000, 1/20000, 1/2000, 1/200$ and $1/20$, respectively. The 
only difference is
that for the type II billiard the escape time $\tau^{\prime}$ is related to 
$\tau$ by $\tau+t_p$, where $t_p$ is the time the trajectory needs to 
travel the two parallel boundaries from the escape hole (see Type II border 
in Fig.~\ref{Model-Border}).
%%%%%%%%%%%%%%%%%%%%%%%%%%%%%%%%%%%%%%%%%%%%%%%%%%%%%%%%%%%%%%%%
%%%%%%%%%%%%%%%%%%%%%%%%%%%%%%%%%%%%%%%%%%%%%%%%%%%%%%%%%%%%%%%%%%%%%
First observation is that for  $R/L=0$, $Q(\tau)$ did not has a power law
(neither an exponential decay). 
This is the integrable case and the maximal ET found is 
$\tau\sim 5.7\times 10^2$. Such maximal time is obtained because we did not
start initial conditions close to the MUPOs ($\theta=0,\pi/2$). Separately
(not shown) we used only initial conditions very close to the MUPOs. We found 
that $Q(\tau)$ 
did not have a finite maximal time, but a power law tail for very long times 
with $\gamma_{esc}\sim1$. This shows that the MUPOs without the dispersing 
component (and no chaotic motion), does not generates the exponent 
$\gamma_{esc}\sim2$. 
This agrees with results obtained \cite{ott04} for some specific initial 
conditions in an open billiard.

For very small rounding effects, $R/L=1/10000$, the qualitative behavior of 
$Q(\tau)$ starts to change  (when compared to the case $R/L=0$) for times 
$\tau\gtrsim 5.7\times 10^2$, 
i.~e.~for those trajectories which stay longer inside the billiard. This 
means that the very few trajectories which collide with the rounded border 
tend to stay longer inside the billiard and also change the qualitative 
behavior of $Q(\tau)$.  The fitted  escape exponent is $\gamma_{esc}\sim 0.6$. 
By increasing the border to $R/L=1/1000$, we observe a power law decay in 
Fig.~\ref{QtR} for times $\tau\gtrsim 5.5\times 10^2$. The escape 
exponent is $\gamma_{esc}\sim 1.3$.  For  $R/L=1/100$ and 
$\tau\gtrsim 3\times 10^{3}$ we obtain  $\gamma_{esc}\sim 1.8$ and
for $R/L=1/10$ and $\tau\gtrsim 1\times 10^{3}$ we obtain
 $\gamma_{esc}\sim 2.1$. From these points is it possible to 
get $\gamma_{esc}\sim 2.5+0.17\ln(R/L)$ and consequently 
we obtain the anomalous exponent $\mu\sim0.5+0.17\ln(R/L)$ for the sticky
region.

The ETs from the long living trajectories present significant characteristics 
of sticky motion for $R/L=1/1000, 1/100, 1/10$. In other words, sticky motion 
and long living trajectories start to occur for {\it very} small rounding 
borders:  around $0.1\%$ from the whole billiard size is sufficient to 
generate the sticky motion and change the output signal. Visually such borders 
are almost negligible. Take for example the border in Fig.~\ref{Model-Border}, 
it has a ratio $R/L\sim1/143$. 

\section{Rounding borders generating the rich dynamics}
\label{dynamics}

\begin{figure}[htb]
 \unitlength 1mm
 \begin{center}
\includegraphics*[width=9cm,angle=0]{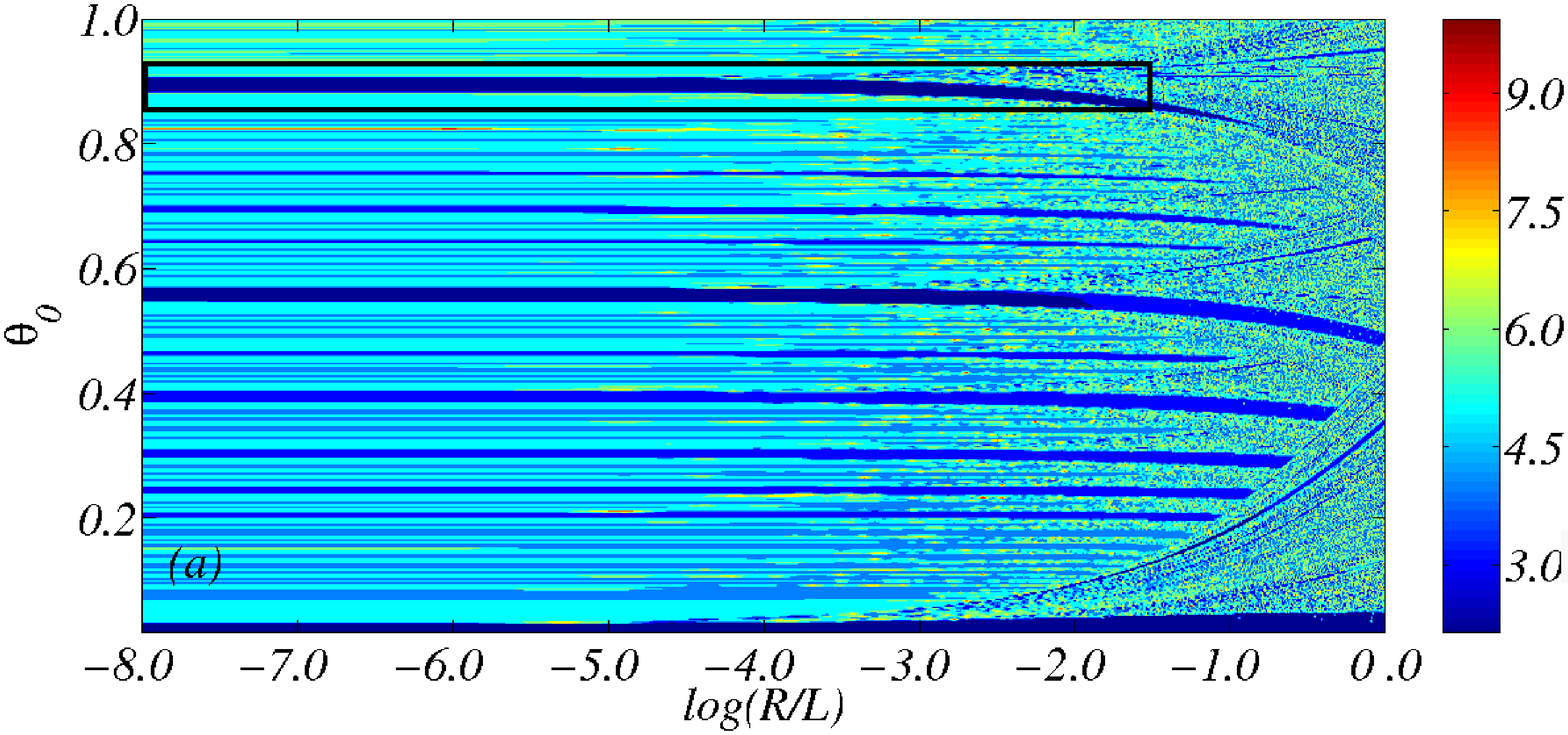}
\includegraphics*[width=9cm,angle=0]{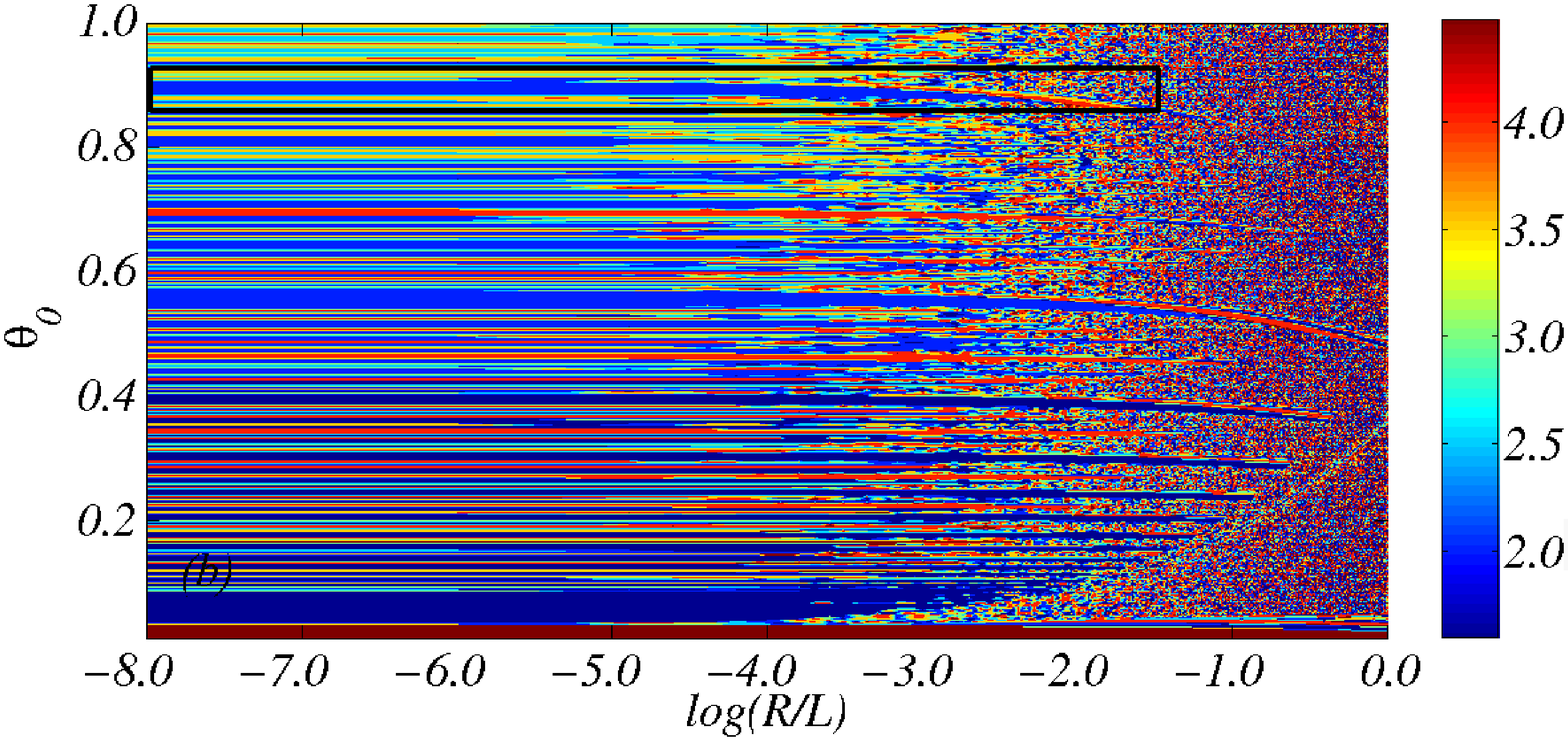}
 \end{center}
\caption{(Color online) (a) Log of the escape time and (b) emission angle 
$\theta_f$ as a function of $\log{(R/L)}$ and $\theta_0$.}
  \label{time}
  \end{figure}

The physics involved in the dynamics becomes more evident when the 
ETs ($\tau$) and escape angles $\theta_f$ are plotted as function of 
the initial incoming angle $\theta_0$ and for different ratios $R/L$. 
These plots are shown
in Fig.~\ref{time} and were generated by using $500\times 500$ points 
in the intervals $0.01\le\theta_0 \le 1.0$ and $0.00355\le R/L\le 1.0$ 
[$-8.0\le \log{(R/L)}\le 0.0$]. We start discussing Fig.~\ref{time}(a),
where each color represents a given value of the ETs written as
$\log{(\tau)}$ (See the colobar on the right). Horizontal stripes with 
different colors are evident for a
significant range of $R/L$ values. Each stripe is defined by a bunch of 
initial conditions which leave to the same ET and consequently have the
same color. For example, for some specific initial angles 
($\theta_0\sim 0.39, 0.56, 0.69, 0.89$) we observe dark blue stripes which 
correspond to very short ETs. For $R/L=0$ these angles can be obtained 
analytically for periodic orbits (for the close billiard) with period $2n$.
They are $\theta_0^{(n)}=\arctan{\left[\frac{D}{2nL}\right]}$, where 
$n=1,2,3,\ldots$, which is the middle value of the main dark blue stripes 
which are born
at $R/L=0$. The corresponding ETs are $t_n=nt$, where $t$ is ETs from the 
case $n=1$, explained below. The shortest ET ($t_0\sim8,0$) occurs for
$\theta_0\sim 0.0$, 
where the particle collides once against the wall in front of the escape hole 
and than leaves the billiard [see trajectory $A_1$ in Fig.~\ref{Model-Border}]. 
The width of the stripes are always related to the aperture $a$ from the hole. 
The next shortest ETs ($t_1=t\sim 12.8$) occurs for $n=1$ at 
$\theta^{(1)}=\arctan{5.0/4.0}\sim 0.89$  where the particle collides 
directly with the edge of the billiard. See trajectory $A_2$ in 
Fig.~\ref{Model-Border}. The third shortest ETs ($\sim 2t$) occurs for 
$\theta^{(2)}=\arctan{5.0/8.0}\sim 0.56$ and is shown by 
trajectory $A_4$  in Fig.~\ref{Model-Border}. We classify these trajectories 
as $A_{2n}$ and, as we will see later, are the starting point from the Arnold 
tongues. In the limit $\theta_0\to \pi/2$ the ETs tend to 
increase since the trajectories are parallel to the escape point. The
ETs statistics for these trajectories will not be considered here, since the
large ET is artificially created due to the location of the escape point in
the horizontal axis.
As $n$ increases the ETs from the trajectories $A_{2n}$ increase, the 
corresponding stripes assume other colors (light blue $\to$ green $\to$ yellow)
[see Fig.~\ref{time}(a)] and their widths decrease. This is the
main behavior of the ETs and stripes close to $R/L\sim 0.0$. 
%%%%%%%%%%%%%%%%%%%%%%%%%%%%%%%%%%%%%%%%%%%%%%%%%%%%%%%%%%%%%%%%%%%%%
%%%%%%%%%%%%%%%%%%%%%%%%%%%%%%%%%%%%%%%%%%%%%%%%%%%%%%%%%%%%%%%%%%%%%

Increasing $R/L$ we observe in Fig.~\ref{time}(a) that the stripes for shorter 
ETs (dark blue) tend to survive longer the bounding border effect, while the 
stripes for larger 
ETs (light blue, green and yellow) tend to be destroyed, or even to mix 
themselves. Notation: we call the dark blue stripes as the {\it main stripes} 
and the other stripes as the {\it secondary stripes}. Therefore, secondary
stripes are always related to intermediate and larger values of ETs. Arnold 
tongues \cite{lichtenberg92} are visible  in Fig.~\ref{time}(a) for 
$\log{(R/L)}\gtrsim-4.0$ and are born around the main stripes with frequencies 
$1/1, 1/2, 1/4, \ldots$. Their relation with the trajectories from 
Fig.~\ref{Model-Border} is: $A_1\to 1/1, \, A_2\to 1/2,\, A_4\to 1/4, \ldots$.
Outside the main stripes the dynamics becomes very rich and complex. Before 
explaining how this occurs, we would like 
to bring to attention the rich dynamics generated by the rounding effects. 

Figure \ref{time}(b) shows $\theta_f$ as a function of $\log{(R/L)}$ and 
$\theta_0$. Each color is now related to one emission angle $\theta_f$ (see 
the colorbar). These emission angles vary between $\theta_f\sim 1.4$ (almost 
horizontally to the left) and  $\theta_f\sim 4.5$ (almost horizontally to the 
right). As in Fig.~\ref{time}(a), also here stripes with different colors 
are evident for a significant range of $R/L$ values. Each stripe is defined 
by a bunch of initial conditions which leave to the same $\theta_f$. In most 
cases these stripes occur for the same values of $\theta_0$ from 
Fig.~\ref{time}(a). However, two stripes with the same color (same ETs) in 
Fig.~\ref{time}(a) have not necessarily the same color (same $\theta_f$) in 
Fig.~\ref{time}(b). In other words, different escape angles can have the
same ETs. As $R/L$ increases more and more, some stripes survive while the 
other ones are destroyed or mixed, as in  Fig.\ref{time}(a). The emission 
angles show a very rich dynamics due to the increasing rounded borders, 
alternating between all possible colors. This will be discussed below in 
more details.

\section{Rounding borders generating backgammon stripes}
\label{backgammon}

\begin{figure}[htb]
 \unitlength 1mm
 \begin{center}
\includegraphics*[width=9cm,angle=0]{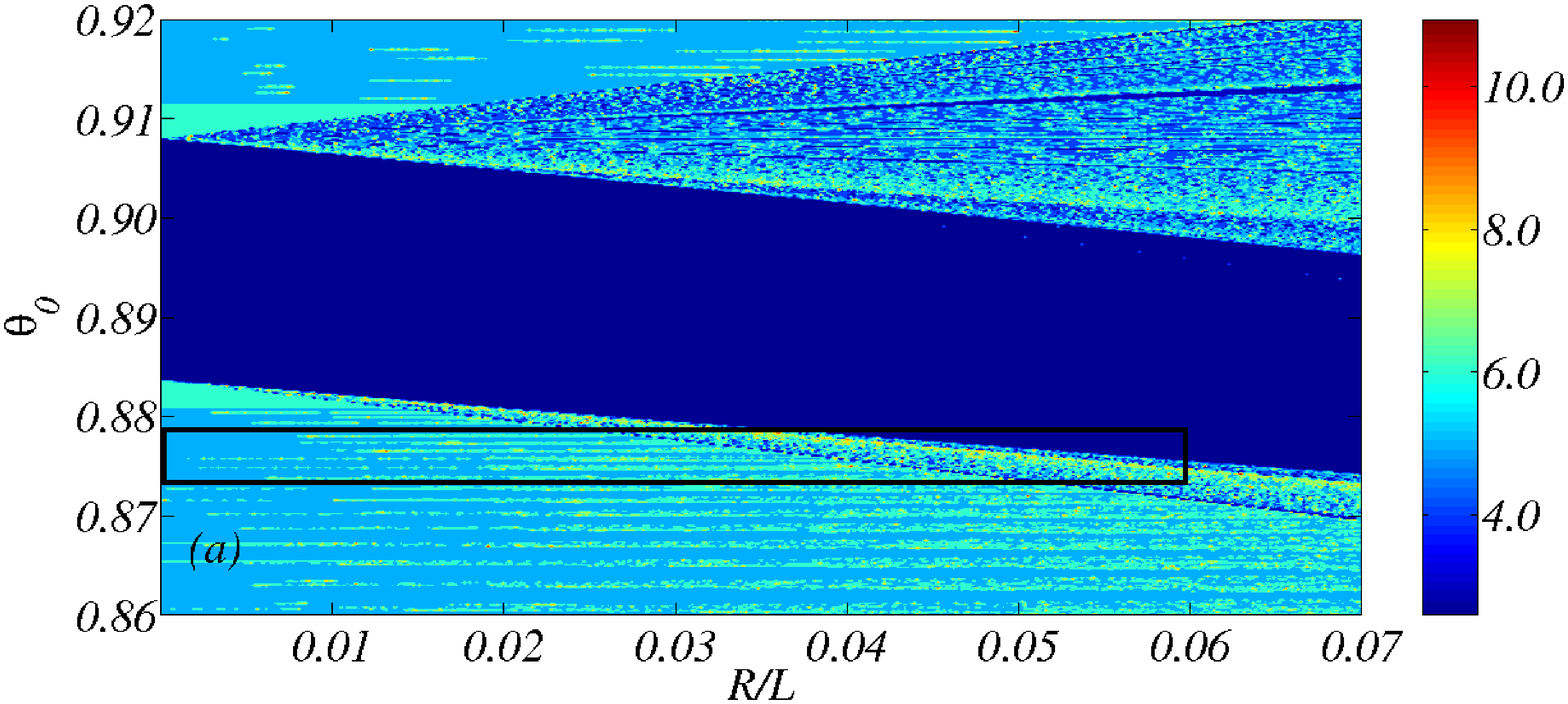}
\includegraphics*[width=9cm,angle=0]{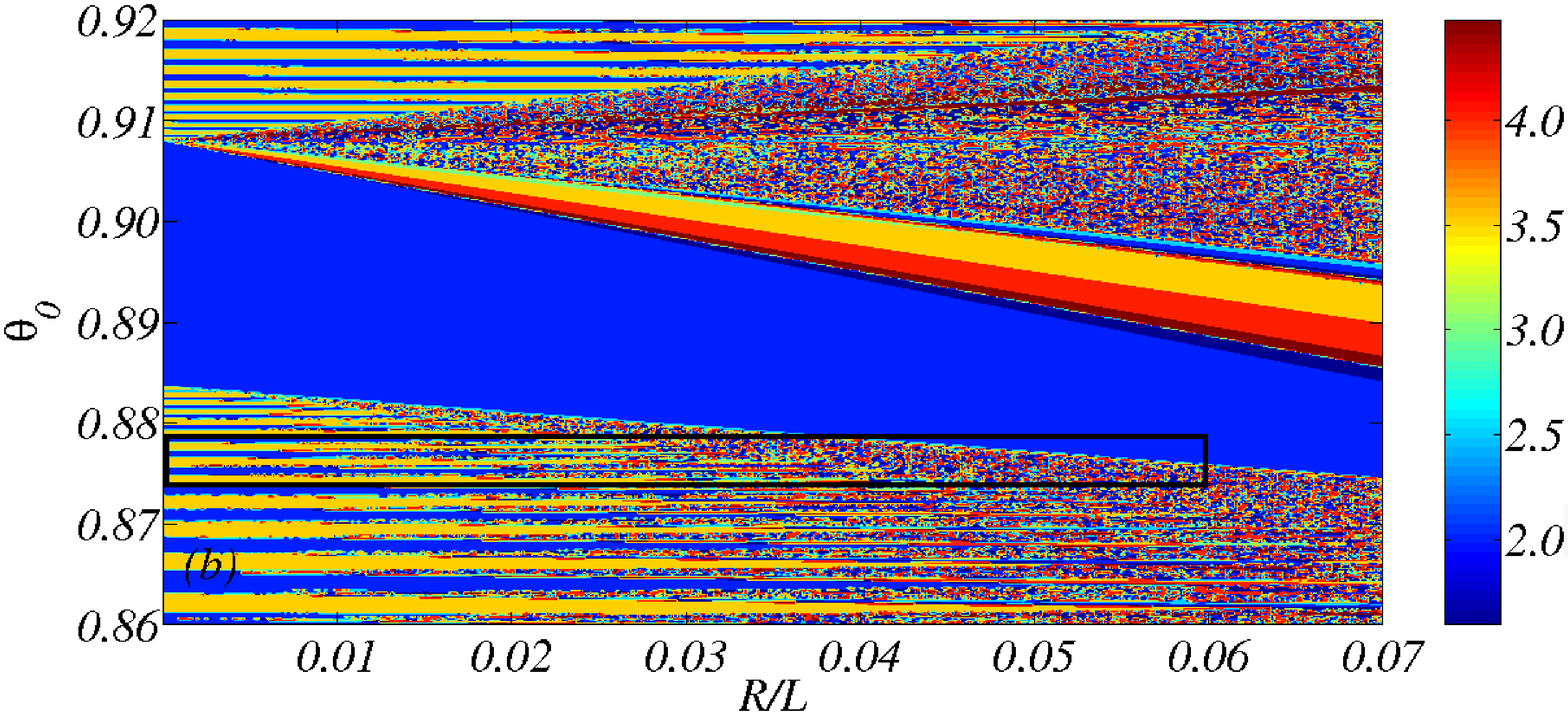}
 \end{center}
\caption{(Color online) Magnifications from the boxes shown 
in Fig.~\ref{time} (a) and (b), respectively.}
  \label{zoom}
  \end{figure}

\begin{figure}[htb]
 \unitlength 1mm
 \begin{center}
\includegraphics*[width=8.5cm,angle=0]{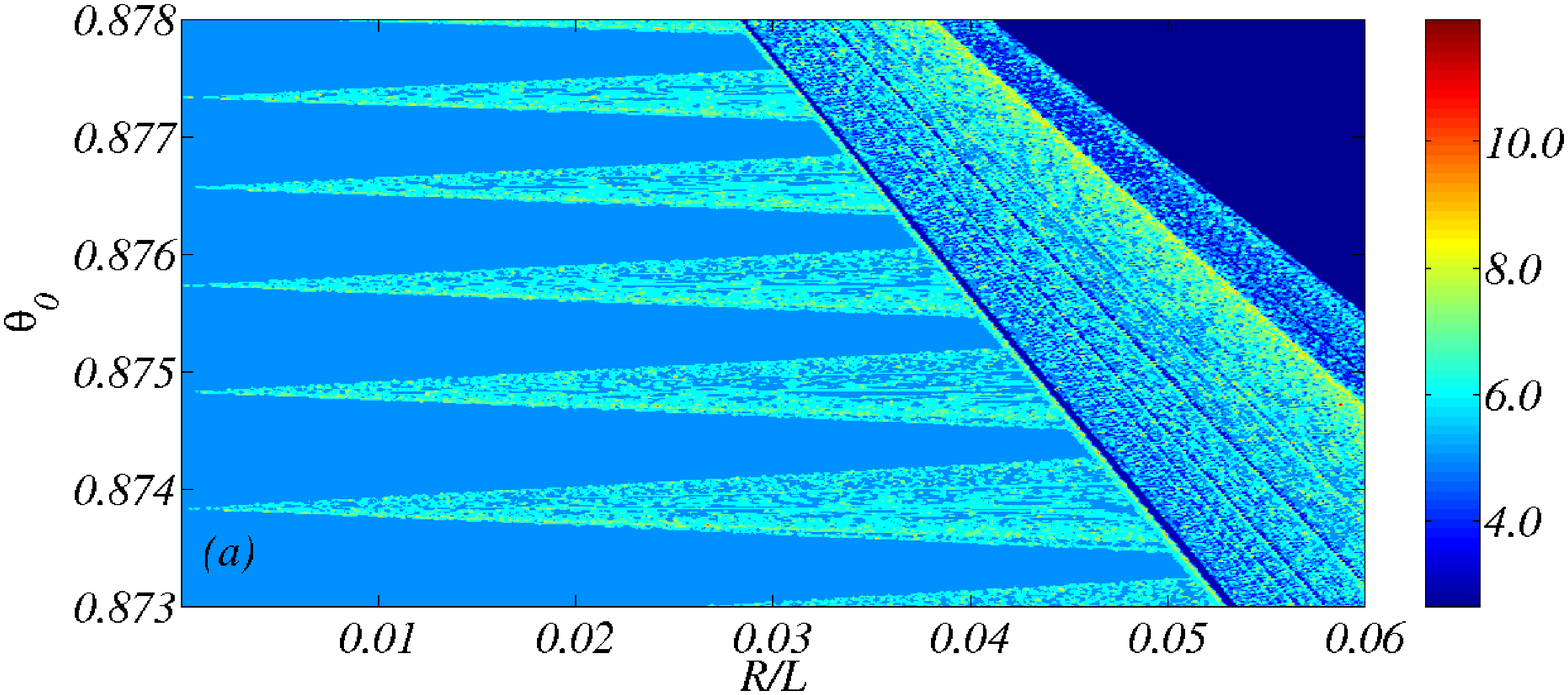}
\includegraphics*[width=8.5cm,angle=0]{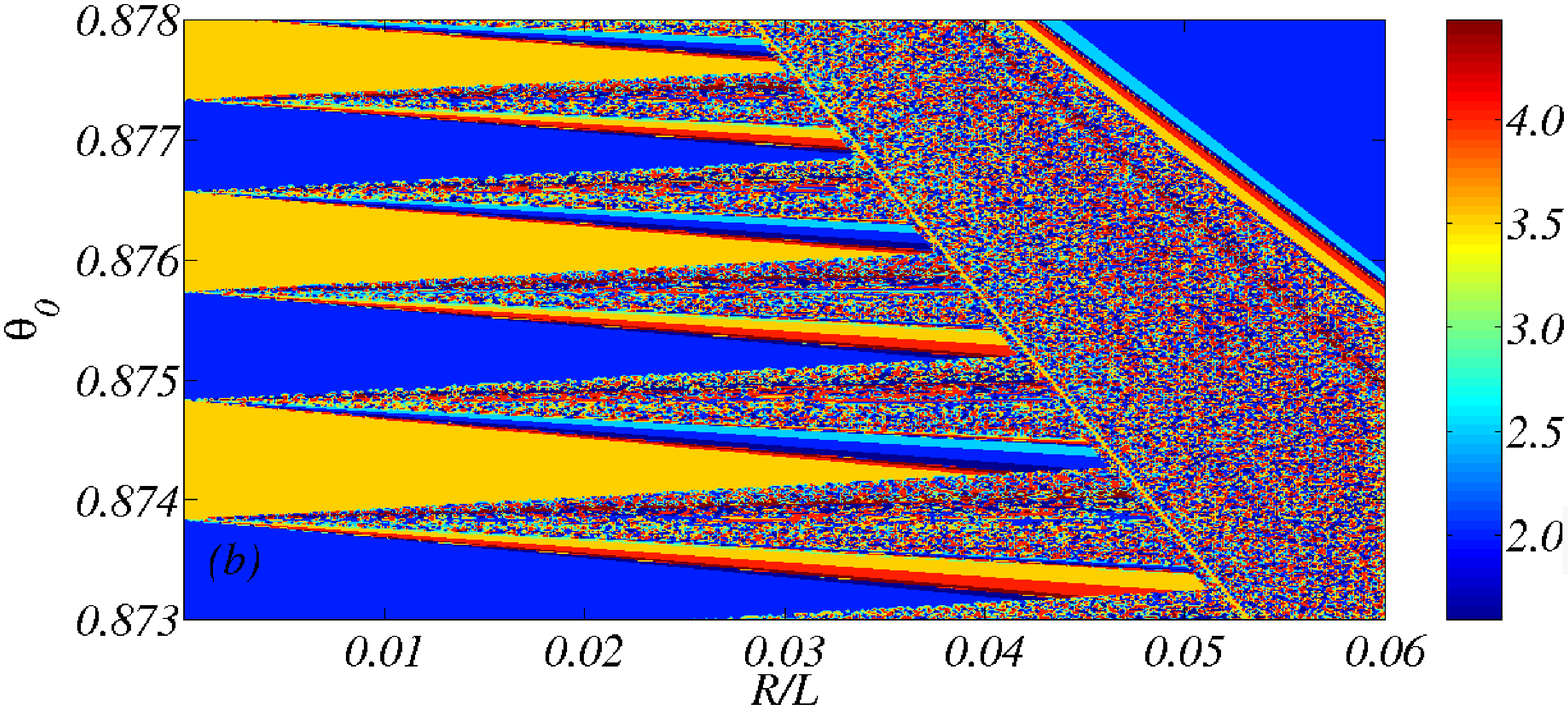}
 \end{center}
\caption{(Color online) Magnifications from the boxes shown 
  in Fig.~\ref{zoom}, respectively.}
  \label{zoom2}
  \end{figure}

In Fig.~\ref{zoom}(a) and (b) we show, respectively, a magnification
for $\log{(\tau)}$ and $\theta_f$ close to the first region. The magnification
is taken around the trajectory $A_2$ (see Fig.~\ref{Model-Border}) which has
a short ETs (main stripe). We observe in Fig.~\ref{zoom}(a) that above the main
stripe the ETs dynamics changes significantly when $R/L$ increases: larger 
(light blue and yellow) and shorter (dark blue) ETs appear inside a secondary 
stripe. This secondary stripe increases linearly its width with $R/L$. For the 
emission angle [see Fig.~\ref{zoom}(b)] we see a very rich dynamics emerging 
inside such a secondary stripe, alternating between all possible colors (all 
emission angles). In addition, below the main stripe a sequence of secondary 
stripes appear in the light blue background, as can observed in the 
magnification shown in Fig.~\ref{zoom2}(a). The width of each stripe in this 
sequence increases with 
$R/L$, remembering stripes from a backgammon board. Notation: secondary stripes 
with increasing width will be referred as ``backgammon stripes''.
Stickiness and the power law behavior observed in Fig.~\ref{QtR} are 
born for initial angles which start inside the  backgammon stripes. These are 
the initial conditions which collide, at least once, with the rounded border.
%%%%%%%%%%%%%%%%%%%%%%%%%%%%%%%%%%%%%%%%%%%%%%%%%%%%%%%%%%%%%%%%%%%%%
%%%%%%%%%%%%%%%%%%%%%%%%%%%%%%%%%%%%%%%%%%%%%%%%%%%%%%%%%%%%%%%%%%%%
The emergence of the power law behavior becomes more evident if we compare 
Fig.~\ref{zoom2}(a) with the emission angle behavior shown in 
Fig.~\ref{zoom2}(b). The light blue background observed in
Fig.~\ref{zoom2}(a), which corresponds to {\it one} ETs, has two colors (blue 
and orange) in Fig.~\ref{zoom2}(b), which correspond to {\it two} emission 
angles
($\sim 2.1$ and $\sim 3.5$). Interesting is that the sequence of backgammon
stripes in Fig.~\ref{zoom2}(a), and the corresponding multicolor backgammon
stripes from Fig.~\ref{zoom2}(b), are born exactly at the boundary between the 
blue and orange escape angles at $R/L\sim 0$. Inside the backgammon stripes 
the range of allowed ETs increases very much. This can be observed by noting 
the increasing number of yellow and red points. The dy\-na\-mics involved in 
the emission angles inside the backgammon stripes is also impressive, showing 
that tiny changes or errors in the initial angle may drastically change the 
emission angle.

The key observation here is that the dynamics {\it inside} the backgammon 
stripes is the consequence of trajectories which  collide with the 
{\it inner part} of the semicircle from the escape point, generating the 
power law behavior for $Q(\tau)$ (see Fig.~\ref{QtR}). Another observation is 
that the location of the backgammon stripes itself is not self-similar but 
{\it inside} the backgammon stripes the self-similar structure is evident.
Many simulations (not shown) were performed to check this statement.
As $R/L$ increases more and more, the self-similar structures increase 
very fast, {\it always} in form of backgammon stripes which emerge at different 
initial angles at $R/L\sim0.0$ (this was checked for many other initial angles).
%%%%%%%%%%%%%%%%%%%%%%%%%%%%%%%%%%%%%%%%%%%%%%%%%%%%%%%%%%%%%%%%%%%%%
%%%%%%%%%%%%%%%%%%%%%%%%%%%%%%%%%%%%%%%%%%%%%%%%%%%%%%%%%%%%%%%%%%%%%
A simple geometrical calculation determines the angle interval 
$[-\theta^*,\theta^*]$ for which at least one collision with the rounded 
border occurs, where 

\begin{equation}
\theta^*=\arctan{\left\{\frac{\sin{\alpha}}{(1-\cos{\alpha})+ a/2R}\right\}},
\nonumber
\end{equation} 
and $\alpha$ is the collision angle of the trajectory with the rounded 
border, defined relative to the center of the semicircle. The interval 
$[-\theta^*,\theta^*]$ defines the border lines of the backgammon stripes.
The larger ETs which appear inside the backgammon stripes start to 
dissapear when the chaotic region is reached close to $R/L\sim 1.0$.
%%%%%%%%%%%%%%%%%%%%%%%%%%%%%%%%%%%%%%%%%%%%%%%%%%%%%%%%%%%%%%%%%%%%%

\section{Conclusions}
\label{conclusions}

Since experiments usually measure a signal coming out from the system of 
interest, the description of open physical devices is of utmost
relevance. For experiments at the frontier of technological limitations, the 
open set (the escape point or hole)
can present irregularities or defects which may induce undesirable and 
intrinsic changes in the outcoming signal. In this work we show that rounded 
borders in the open set generate a rich dynamics 
in the integrable rectangular billiard. The escape times statistics for 
the long-lived trajectories present characteristic of sticky motion when the
rounded border represents around $0.1\%\to 10\%$ from the whole 
billiard size. In this sticky region the escape times decay 
exponent $\gamma_{esc}$ increases with $R/L$ and approaches
$\gamma_{esc}\sim2.0$ for $R/L=1/10$ ($R$ is the radius of the rounded border 
and $L$ is the height of the billiard). Emission angles and 
escape times show a self-similar structure only for initial angles inside 
backgammon like stripes which are born at the integrable 
case $R/L=0$. Trajectories which start inside the  backgammon stripes
will collide, at least once, with the rounded border, generating the power
law behavior. As $R/L$ increases more 
and more, different backgammon stripes start to overlap and the exponent 
$\gamma_{esc}\sim2.0$ is obtained.
For $R/L=1.0$ the exponential decay for the escape times is reached, which 
corresponds to the chaotic motion. 

From the nonlinear perspective our results are impressive, showing that a 
very rich dynamics and stickiness comes alone from tiny border effects 
($\sim 0.1\%$) in integrable billiards. Such effects, including the Arnold 
tongues, should be 
visible directly in experiments with open integrable devices, open systems
with leaks (see \cite{altmann09} and therein cited references) and also
in problems related to conduction fluctuations in semiconductors 
\cite{geisel07,marlow06}, particles transport in nanostructures and cold atoms 
in open optical billiards \cite{kaplan09}.

Instead of rounded borders, other shapes for the open sets could be used. All
of them should generate sticky motion in integrable billiards, strongly 
affecting the outcoming signal. Our results will change quantitatively when 
the aperture $a$ increases, however, in such cases 
long-lived trajectories will disappear and the statistical analysis for the 
escape times is not adequate anymore. In this work we considered an integrable 
billiard as the starting point. We also 
performed extensive numerical si\-mu\-la\-tions to study rounded border 
effects in the stadium billiard, which is already chaotic without border 
effects ($R/L=0$). For all values of $R/L$ we found that the escape times 
statistic has an exponential decay, expected for the chaotic behavior.

\begin{acknowledgments}
The authors thank CNPq, CAPES and FINEP (under project CT-INFRA/UFPR) for 
partial financial support. They also thank E.~Altmann and C.~Manchein for 
helpfull discussions.
\end{acknowledgments}

%\bibliography{references}

\begin{thebibliography}{34}
\expandafter\ifx\csname natexlab\endcsname\relax\def\natexlab#1{#1}\fi
\expandafter\ifx\csname bibnamefont\endcsname\relax
  \def\bibnamefont#1{#1}\fi
\expandafter\ifx\csname bibfnamefont\endcsname\relax
  \def\bibfnamefont#1{#1}\fi
\expandafter\ifx\csname citenamefont\endcsname\relax
  \def\citenamefont#1{#1}\fi
\expandafter\ifx\csname url\endcsname\relax
  \def\url#1{\texttt{#1}}\fi
\expandafter\ifx\csname urlprefix\endcsname\relax\def\urlprefix{URL }\fi
\providecommand{\bibinfo}[2]{#2}
\providecommand{\eprint}[2][]{\url{#2}}

\bibitem[{\citenamefont{Libisch et~al.}(2009)\citenamefont{Libisch, Stampfer,
  and Burgd\"orfer}}]{libisch09}
\bibinfo{author}{\bibfnamefont{F.}~\bibnamefont{Libisch}},
  \bibinfo{author}{\bibfnamefont{C.}~\bibnamefont{Stampfer}}, \bibnamefont{and}
  \bibinfo{author}{\bibfnamefont{J.}~\bibnamefont{Burgd\"orfer}},
  \bibinfo{journal}{Phys.~Rev.~B} \textbf{\bibinfo{volume}{79}},
  \bibinfo{pages}{115423} (\bibinfo{year}{2009}).

\bibitem[{\citenamefont{Bogomolny et~al.}(2009)\citenamefont{Bogomolny, Dennis,
  and Dubertrand}}]{bogo09}
\bibinfo{author}{\bibfnamefont{E.}~\bibnamefont{Bogomolny}},
  \bibinfo{author}{\bibfnamefont{M.~R.} \bibnamefont{Dennis}},
  \bibnamefont{and}
  \bibinfo{author}{\bibfnamefont{D.}~\bibnamefont{Dubertrand}},
  \bibinfo{journal}{J.~Phys.~A} \textbf{\bibinfo{volume}{41}},
  \bibinfo{pages}{335102} (\bibinfo{year}{2009}).

\bibitem[{\citenamefont{Berry and Dennis}(2008)}]{dennis08}
\bibinfo{author}{\bibfnamefont{M.~V.} \bibnamefont{Berry}} \bibnamefont{and}
  \bibinfo{author}{\bibfnamefont{M.~R.} \bibnamefont{Dennis}},
  \bibinfo{journal}{J.~Phys.~A} \textbf{\bibinfo{volume}{41}},
  \bibinfo{pages}{135203} (\bibinfo{year}{2008}).

\bibitem[{\citenamefont{Oliveira et~al.}(2008)\citenamefont{Oliveira, Manchein,
  and Beims}}]{hercules08}
\bibinfo{author}{\bibfnamefont{H.~A.} \bibnamefont{Oliveira}},
  \bibinfo{author}{\bibfnamefont{C.}~\bibnamefont{Manchein}}, \bibnamefont{and}
  \bibinfo{author}{\bibfnamefont{M.~W.} \bibnamefont{Beims}},
  \bibinfo{journal}{Phys.~Rev.~E} \textbf{\bibinfo{volume}{78}},
  \bibinfo{pages}{046208} (\bibinfo{year}{2008}).

\bibitem[{\citenamefont{Beims et~al.}(2007)\citenamefont{Beims, Manchein, and
  Rost}}]{cesar1}
\bibinfo{author}{\bibfnamefont{M.~W.} \bibnamefont{Beims}},
  \bibinfo{author}{\bibfnamefont{C.}~\bibnamefont{Manchein}}, \bibnamefont{and}
  \bibinfo{author}{\bibfnamefont{J.~M.} \bibnamefont{Rost}},
  \bibinfo{journal}{Phys. Rev. E} \textbf{\bibinfo{volume}{76}},
  \bibinfo{pages}{056203} (\bibinfo{year}{2007}).

\bibitem[{\citenamefont{Wiersig}(2003)}]{wiersig03}
\bibinfo{author}{\bibfnamefont{J.}~\bibnamefont{Wiersig}},
  \bibinfo{journal}{Phys.~Rev.~A} \textbf{\bibinfo{volume}{67}},
  \bibinfo{pages}{023807} (\bibinfo{year}{2003}).

\bibitem[{\citenamefont{Vaa et~al.}(2005)\citenamefont{Vaa, Koch, and
  Bl\"umel}}]{blumel05}
\bibinfo{author}{\bibfnamefont{C.}~\bibnamefont{Vaa}},
  \bibinfo{author}{\bibfnamefont{P.~M.} \bibnamefont{Koch}}, \bibnamefont{and}
  \bibinfo{author}{\bibfnamefont{R.}~\bibnamefont{Bl\"umel}},
  \bibinfo{journal}{Phys.~Rev.~E} \textbf{\bibinfo{volume}{72}},
  \bibinfo{pages}{056211} (\bibinfo{year}{2005}).

\bibitem[{\citenamefont{Dubertrand et~al.}(2008)\citenamefont{Dubertrand,
  Bogomolny, Djellali, Lebental, and Schmit}}]{bogo08}
\bibinfo{author}{\bibfnamefont{R.}~\bibnamefont{Dubertrand}},
  \bibinfo{author}{\bibfnamefont{E.}~\bibnamefont{Bogomolny}},
  \bibinfo{author}{\bibfnamefont{N.}~\bibnamefont{Djellali}},
  \bibinfo{author}{\bibfnamefont{M.}~\bibnamefont{Lebental}}, \bibnamefont{and}
  \bibinfo{author}{\bibfnamefont{C.}~\bibnamefont{Schmit}},
  \bibinfo{journal}{Phys.~Rev.~A} \textbf{\bibinfo{volume}{77}},
  \bibinfo{pages}{013804} (\bibinfo{year}{2008}).

\bibitem[{\citenamefont{Bruus and Whelan}(1996)}]{whelan96}
\bibinfo{author}{\bibfnamefont{H.}~\bibnamefont{Bruus}} \bibnamefont{and}
  \bibinfo{author}{\bibfnamefont{N.~D.} \bibnamefont{Whelan}},
  \bibinfo{journal}{Nonlinearity} \textbf{\bibinfo{volume}{9}},
  \bibinfo{pages}{1023} (\bibinfo{year}{1996}).

\bibitem[{\citenamefont{Alonso and Gaspard}(1994)}]{gaspard94}
\bibinfo{author}{\bibfnamefont{D.}~\bibnamefont{Alonso}} \bibnamefont{and}
  \bibinfo{author}{\bibfnamefont{P.}~\bibnamefont{Gaspard}},
  \bibinfo{journal}{J.~Phys.~A} \textbf{\bibinfo{volume}{27}},
  \bibinfo{pages}{1599} (\bibinfo{year}{1994}).

\bibitem[{\citenamefont{Marlow et~al.}(2006)\citenamefont{Marlow, Taylor,
  Martin, B.C.Scannell, Linke, Fairbanks, Hall, Shorubalko, Fromhold, Brown
  et~al.}}]{marlow06}
\bibinfo{author}{\bibfnamefont{C.}~\bibnamefont{Marlow}},
  \bibinfo{author}{\bibfnamefont{R.}~\bibnamefont{Taylor}},
  \bibinfo{author}{\bibfnamefont{T.}~\bibnamefont{Martin}},
  \bibinfo{author}{\bibnamefont{B.C.Scannell}},
  \bibinfo{author}{\bibfnamefont{H.}~\bibnamefont{Linke}},
  \bibinfo{author}{\bibfnamefont{M.}~\bibnamefont{Fairbanks}},
  \bibinfo{author}{\bibfnamefont{G.}~\bibnamefont{Hall}},
  \bibinfo{author}{\bibfnamefont{I.}~\bibnamefont{Shorubalko}},
  \bibinfo{author}{\bibfnamefont{L.~T.} \bibnamefont{Fromhold}},
  \bibinfo{author}{\bibfnamefont{C.}~\bibnamefont{Brown}},
  \bibnamefont{et~al.}, \bibinfo{journal}{Phys.~Rev.~B}
  \textbf{\bibinfo{volume}{73}}, \bibinfo{pages}{195318}
  (\bibinfo{year}{2006}).

\bibitem[{\citenamefont{Sachrajda et~al.}(1998)\citenamefont{Sachrajda,
  Ketzmerick, Gould, Feng, Kelly, Delage, and Wasilewski}}]{sachrajda98}
\bibinfo{author}{\bibfnamefont{A.~S.} \bibnamefont{Sachrajda}},
  \bibinfo{author}{\bibfnamefont{R.}~\bibnamefont{Ketzmerick}},
  \bibinfo{author}{\bibfnamefont{C.}~\bibnamefont{Gould}},
  \bibinfo{author}{\bibfnamefont{Y.}~\bibnamefont{Feng}},
  \bibinfo{author}{\bibfnamefont{P.~J.} \bibnamefont{Kelly}},
  \bibinfo{author}{\bibfnamefont{A.}~\bibnamefont{Delage}}, \bibnamefont{and}
  \bibinfo{author}{\bibfnamefont{Z.}~\bibnamefont{Wasilewski}},
  \bibinfo{journal}{Phys. Rev. Lett.} \textbf{\bibinfo{volume}{80}},
  \bibinfo{pages}{1948} (\bibinfo{year}{1998}).

\bibitem[{\citenamefont{Takagaki et~al.}(2000)\citenamefont{Takagaki, Ploog,
  Lin, Aoki, and Ochiai}}]{takagaki00}
\bibinfo{author}{\bibfnamefont{Y.}~\bibnamefont{Takagaki}},
  \bibinfo{author}{\bibfnamefont{K.~H.} \bibnamefont{Ploog}},
  \bibinfo{author}{\bibfnamefont{L.-H.} \bibnamefont{Lin}},
  \bibinfo{author}{\bibfnamefont{N.}~\bibnamefont{Aoki}}, \bibnamefont{and}
  \bibinfo{author}{\bibfnamefont{Y.}~\bibnamefont{Ochiai}},
  \bibinfo{journal}{Phys.~Rev.~B} \textbf{\bibinfo{volume}{62}},
  \bibinfo{pages}{10255} (\bibinfo{year}{2000}).

\bibitem[{\citenamefont{Hennig et~al.}(2007)\citenamefont{Hennig, Fleischmann,
  Hufnagel, and Geisel}}]{geisel07}
\bibinfo{author}{\bibfnamefont{H.}~\bibnamefont{Hennig}},
  \bibinfo{author}{\bibfnamefont{R.}~\bibnamefont{Fleischmann}},
  \bibinfo{author}{\bibfnamefont{L.}~\bibnamefont{Hufnagel}}, \bibnamefont{and}
  \bibinfo{author}{\bibfnamefont{T.}~\bibnamefont{Geisel}},
  \bibinfo{journal}{Phys.~Rev.~E} \textbf{\bibinfo{volume}{76}},
  \bibinfo{pages}{015202} (\bibinfo{year}{2007}).

\bibitem[{\citenamefont{Kaplan et~al.}(2001)\citenamefont{Kaplan, Friedman,
  Andersen, and Davidson}}]{kaplan09}
\bibinfo{author}{\bibfnamefont{A.}~\bibnamefont{Kaplan}},
  \bibinfo{author}{\bibfnamefont{N.}~\bibnamefont{Friedman}},
  \bibinfo{author}{\bibfnamefont{M.}~\bibnamefont{Andersen}}, \bibnamefont{and}
  \bibinfo{author}{\bibfnamefont{N.}~\bibnamefont{Davidson}},
  \bibinfo{journal}{Phys.~Rev.~Lett.} \textbf{\bibinfo{volume}{87}},
  \bibinfo{pages}{274101} (\bibinfo{year}{2001}).

\bibitem[{\citenamefont{Altmann and T\'el}(2009)}]{altmann09}
\bibinfo{author}{\bibfnamefont{E.~G.} \bibnamefont{Altmann}} \bibnamefont{and}
  \bibinfo{author}{\bibfnamefont{T.}~\bibnamefont{T\'el}},
  \bibinfo{journal}{Phys.~Rev.~E} \textbf{\bibinfo{volume}{79}},
  \bibinfo{pages}{016204} (\bibinfo{year}{2009}).

\bibitem[{\citenamefont{Dettmann and Georgiou}(2009)}]{dettmann09}
\bibinfo{author}{\bibfnamefont{C.~P.} \bibnamefont{Dettmann}} \bibnamefont{and}
  \bibinfo{author}{\bibfnamefont{O.}~\bibnamefont{Georgiou}},
  \bibinfo{journal}{Physica D} \textbf{\bibinfo{volume}{238}},
  \bibinfo{pages}{2395} (\bibinfo{year}{2009}).

\bibitem[{\citenamefont{Bunimovich and Dettmann}(2007)}]{buni07}
\bibinfo{author}{\bibfnamefont{L.~A.} \bibnamefont{Bunimovich}}
  \bibnamefont{and} \bibinfo{author}{\bibfnamefont{C.~P.}
  \bibnamefont{Dettmann}}, \bibinfo{journal}{Europhys.~}
  \textbf{\bibinfo{volume}{80}}, \bibinfo{pages}{40001} (\bibinfo{year}{2007}).

\bibitem[{\citenamefont{Armstead et~al.}(2004)\citenamefont{Armstead, Hunt, and
  Ott}}]{ott04}
\bibinfo{author}{\bibfnamefont{D.~N.} \bibnamefont{Armstead}},
  \bibinfo{author}{\bibfnamefont{B.~R.} \bibnamefont{Hunt}}, \bibnamefont{and}
  \bibinfo{author}{\bibfnamefont{E.}~\bibnamefont{Ott}},
  \bibinfo{journal}{Physica D} \textbf{\bibinfo{volume}{193}},
  \bibinfo{pages}{96} (\bibinfo{year}{2004}).

\bibitem[{\citenamefont{Zaslavski}(2002)}]{zas02}
\bibinfo{author}{\bibfnamefont{G.~M.} \bibnamefont{Zaslavski}},
  \bibinfo{journal}{Phys. Rep.} \textbf{\bibinfo{volume}{371}},
  \bibinfo{pages}{461} (\bibinfo{year}{2002}).

\bibitem[{\citenamefont{Tomsovic and Lakshminarayan}(2007)}]{steven07}
\bibinfo{author}{\bibfnamefont{S.}~\bibnamefont{Tomsovic}} \bibnamefont{and}
  \bibinfo{author}{\bibfnamefont{A.}~\bibnamefont{Lakshminarayan}},
  \bibinfo{journal}{Phys.~Rev.~E} \textbf{\bibinfo{volume}{76}},
  \bibinfo{pages}{036207} (\bibinfo{year}{2007}).

\bibitem[{\citenamefont{Manchein et~al.}(2009)\citenamefont{Manchein, Beims,
  and Rost}}]{cesarPRE09}
\bibinfo{author}{\bibfnamefont{C.}~\bibnamefont{Manchein}},
  \bibinfo{author}{\bibfnamefont{M.~W.} \bibnamefont{Beims}}, \bibnamefont{and}
  \bibinfo{author}{\bibfnamefont{J.~M.} \bibnamefont{Rost}},
  \bibinfo{journal}{arXiv:0907.4181}  (\bibinfo{year}{2009}).

\bibitem[{\citenamefont{Artuso and Manchein}(2009)}]{manchein09}
\bibinfo{author}{\bibfnamefont{R.}~\bibnamefont{Artuso}} \bibnamefont{and}
  \bibinfo{author}{\bibfnamefont{C.}~\bibnamefont{Manchein}},
  \bibinfo{journal}{Phys.~Rev.~E} \textbf{\bibinfo{volume}{80}},
  \bibinfo{pages}{036210} (\bibinfo{year}{2009}).

\bibitem[{\citenamefont{Mitchell and Ilan}(2009)}]{mitchell09}
\bibinfo{author}{\bibfnamefont{K.~A.} \bibnamefont{Mitchell}} \bibnamefont{and}
  \bibinfo{author}{\bibfnamefont{B.}~\bibnamefont{Ilan}},
  \bibinfo{journal}{Phys.~Rev.~A} \textbf{\bibinfo{volume}{80}},
  \bibinfo{pages}{043406} (\bibinfo{year}{2009}).

\bibitem[{\citenamefont{Mitchell and Steck}(2007)}]{mitchell07}
\bibinfo{author}{\bibfnamefont{K.~A.} \bibnamefont{Mitchell}} \bibnamefont{and}
  \bibinfo{author}{\bibfnamefont{D.~A.} \bibnamefont{Steck}},
  \bibinfo{journal}{Phys.~Rev.~A} \textbf{\bibinfo{volume}{76}},
  \bibinfo{pages}{031403} (\bibinfo{year}{2007}).

\bibitem[{\citenamefont{Hansen et~al.}(2006)\citenamefont{Hansen, Mitchell, and
  Delos}}]{delos06}
\bibinfo{author}{\bibfnamefont{P.}~\bibnamefont{Hansen}},
  \bibinfo{author}{\bibfnamefont{K.~A.} \bibnamefont{Mitchell}},
  \bibnamefont{and} \bibinfo{author}{\bibfnamefont{J.~B.} \bibnamefont{Delos}},
  \bibinfo{journal}{Phys.~Rev.~E} \textbf{\bibinfo{volume}{73}},
  \bibinfo{pages}{066226} (\bibinfo{year}{2006}).

\bibitem[{\citenamefont{Ketzmerick}(1996)}]{ketzmerick96}
\bibinfo{author}{\bibfnamefont{R.}~\bibnamefont{Ketzmerick}},
  \bibinfo{journal}{Phys.~Rev.~B} \textbf{\bibinfo{volume}{54}},
  \bibinfo{pages}{10841} (\bibinfo{year}{1996}).

\bibitem[{\citenamefont{Altmann et~al.}(2005)\citenamefont{Altmann, Motter, and
  Kantz}}]{altmann05}
\bibinfo{author}{\bibfnamefont{E.~G.} \bibnamefont{Altmann}},
  \bibinfo{author}{\bibfnamefont{A.}~\bibnamefont{Motter}}, \bibnamefont{and}
  \bibinfo{author}{\bibfnamefont{H.}~\bibnamefont{Kantz}},
  \bibinfo{journal}{CHAOS} \textbf{\bibinfo{volume}{15}},
  \bibinfo{pages}{033105} (\bibinfo{year}{2005}).

\bibitem[{\citenamefont{Meiss}(1997)}]{meiss97}
\bibinfo{author}{\bibfnamefont{J.~D.} \bibnamefont{Meiss}},
  \bibinfo{journal}{Chaos} \textbf{\bibinfo{volume}{7}}, \bibinfo{pages}{139}
  (\bibinfo{year}{1997}).

\bibitem[{\citenamefont{Zumofen and Klafter}(1999)}]{klafter99}
\bibinfo{author}{\bibfnamefont{G.}~\bibnamefont{Zumofen}} \bibnamefont{and}
  \bibinfo{author}{\bibfnamefont{J.}~\bibnamefont{Klafter}},
  \bibinfo{journal}{Phys.~Rev.~E} \textbf{\bibinfo{volume}{59}},
  \bibinfo{pages}{3756} (\bibinfo{year}{1999}).

\bibitem[{\citenamefont{Sinai}(1968)}]{sinai}
\bibinfo{author}{\bibfnamefont{Y.~G.} \bibnamefont{Sinai}},
  \bibinfo{journal}{Func. Anal. Appl.} \textbf{\bibinfo{volume}{2}},
  \bibinfo{pages}{61} (\bibinfo{year}{1968}).

\bibitem[{\citenamefont{Bunimovich}(1979)}]{buni79}
\bibinfo{author}{\bibfnamefont{L.~A.} \bibnamefont{Bunimovich}},
  \bibinfo{journal}{Math.~Phys.~} \textbf{\bibinfo{volume}{65}},
  \bibinfo{pages}{295} (\bibinfo{year}{1979}).

\bibitem[{\citenamefont{Gaspard and Dorfman}(1995)}]{dorfman95}
\bibinfo{author}{\bibfnamefont{P.}~\bibnamefont{Gaspard}} \bibnamefont{and}
  \bibinfo{author}{\bibfnamefont{J.~R.} \bibnamefont{Dorfman}},
  \bibinfo{journal}{Phys.~Rev.~E} \textbf{\bibinfo{volume}{52}},
  \bibinfo{pages}{3525} (\bibinfo{year}{1995}).

\bibitem[{\citenamefont{Lichtenberg and Lieberman}(1992)}]{lichtenberg92}
\bibinfo{author}{\bibfnamefont{A.~J.} \bibnamefont{Lichtenberg}}
  \bibnamefont{and} \bibinfo{author}{\bibfnamefont{M.~A.}
  \bibnamefont{Lieberman}}, \emph{\bibinfo{title}{Regular and Chaotic
  Dynamics}} (\bibinfo{publisher}{Springer-Verlag}, \bibinfo{year}{1992}).

\end{thebibliography}

\end{document}